# Nonlinear optics of photonic hyper-crystals: optical limiting and hyper-computing

IGOR I. SMOLYANINOV

*Department of Electrical and Computer Engineering, University of Maryland, College Park, MD 20742, USA*
*smoly@umd.edu*

**Abstract:** Photonic hyper-crystals combine the most interesting features of hyperbolic metamaterials and photonic crystals. Since the dispersion law of extraordinary photons in hyperbolic metamaterials does not exhibit the usual diffraction limit, photonic hyper-crystals exhibit light localization on deep subwavelength scales, leading to considerable enhancement of nonlinear photon-photon interaction. Therefore, similar to their conventional photonic crystal counterparts, nonlinear photonic hyper-crystals appear to be very promising in optical limiting and optical computing applications. Nonlinear optics of photonic hyper-crystals may be formulated in such a way that one of the spatial coordinates would play a role of effective time in a 2+1 dimensional "optical spacetime" describing light propagation in the hyper-crystal. Mapping the conventional optical computing onto nonlinear optics of photonic hyper-crystals results in a "hyper-computing" scheme, which may considerably accelerate computation time.

## 1. Introduction

Finding an efficient way to make photons interact with each other remains a major challenge in optical limiting and optical computing applications [1]. A considerable recent progress in this area of research has been achieved by using photonic crystals [2]. A photonic crystal-based nanocavity design enables an ultra-small mode volume, so that considerable nonlinearities may be observed even at single-photon levels. This recent success motivates us to look at the nonlinear optics of photonic hyper-crystals [3], which combine the most interesting features of hyperbolic metamaterials and photonic crystals. Photonic hyper-crystals are formed by periodic modulation of hyperbolic metamaterial properties on a scale $L$, which is much smaller than the free space light wavelength $\lambda$. Since the dispersion law of extraordinary photons in hyperbolic metamaterials does not exhibit the conventional diffraction limit, such modulation would lead to Bragg scattering of extraordinary photons and formation of photonic band structure no matter how small $L$ is [4]. Therefore, photonic hyper-crystals exhibit light localization on deep subwavelength scales, which far exceeds the demonstrated mode volume reduction in photonic crystals. It was suggested that such strong field localization would lead to considerable enhancement of nonlinear optical effects in photonic hyper-crystals [3]. In this paper we will demonstrate that the level of nonlinearities achievable in photonic hyper-crystals indeed looks very promising for optical limiting and optical computing applications.

Our goal is to develop insights into classical and quantum nonlinear optics of photonic hyper-crystals and to suggest promising optical limiting and computing geometries, which would make use of the strong nonlinearities mentioned above. Since nonlinear optics of hyperbolic metamaterials finds natural interpretation in terms of analog gravity in an effective 2+1 dimensional "optical spacetime" describing light propagation in the metamaterial [5], it is not surprising that a similar language is useful in describing the nonlinear optics of photonic hyper-crystals. Moreover, periodic modulation of hyperbolic metamaterial properties, which is necessary for hyper-crystal formation, may involve creation of hyperbolic metamaterial interfaces with conventional dielectrics. As demonstrated in [6], such interfaces often behave similar to various horizons in Minkowski space-time. Very strong and tightly localized optical field divergences near the apparent horizons at the hyperbolic metamaterial resonator boundaries ensure strong nonlinear photon-photon interaction in such geometry. We should also mention that recent studies (see for example [7]) demonstrated that the issue of losses in hyperbolic metamaterials may be managed by using gain media as a dielectric component of the metamaterial. Thus, both hyperbolic metamaterials and photonic hyper-crystals may be made lossless (and even active) in a narrow frequency range. Therefore, the use of hyperbolic materials in optical limiting and computing applications is justifiable.

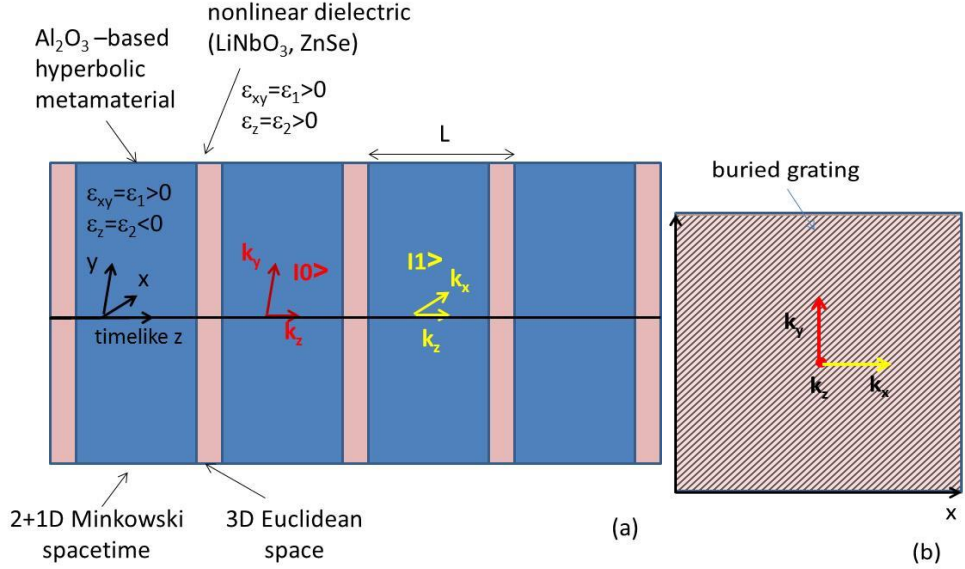

Fig. 1. (a) Schematic geometry of a photonic hyper-crystal based on hyperbolic material resonators separated by thin nonlinear dielectric layers. The optical bit states in this geometry may be encoded via the kxy component of the photon k vector, as indicated by the red and yellow arrows. Effective horizons leading to field divergences occur at hyperbolic/dielectric interfaces, thus enabling nonlinear photon-photon interaction. (b) Side view of the nonlinear dielectric layer having a "buried grating". The grating, which can be switched on and off mixes the optical bit states.

This paper is organized as follows. In Section 2 we introduce the photonic hyper-crystal geometry (Fig.1) of particular interest to optical limiting and optical computing applications, which is made of periodic arrangement of hyperbolic resonators [8]. Linear optical properties of such a photonic hyper-crystal will be described in terms of an effective optical space-time inside an individual resonator. Mutual coupling of these resonators and formation of Bloch waves will be considered as a function of height of the effective potential walls at the resonator boundaries (which look like effective horizons). Section 3 will address nonlinear optical interaction of photons in such a hyperbolic metamaterial resonator array. Section 4 will address potential applications of nonlinear photonic hyper-crystals in optical limiting and optical computing devices. It will be demonstrated that in computing applications optical "0" and "1" bits may be encoded either via light intensity, or via the orthogonal $k_{xy}$ directions in the individual resonators, as illustrated in Fig.1. It appears that presence or absence of photons in a resonator may considerably alter light reflection and transmission at the resonator boundaries, and therefore change nonlinear optical interaction of optical bits. Since one of the spatial coordinates plays the role of effective time in this geometry, the periodic resonator boundaries may play the role of a computer "clock". Such an optical "hyper-computing" scheme, which maps the computer clock cycle onto the periodic modulation of the photonic hyper-crystal, may considerably accelerate computation time, leading to potential advantages over the conventional optical computing schemes. The paper will be concluded by a brief summary of obtained results.

## 2. Linear optical properties of hyperbolic resonator-based photonic hyper-crystals

Let us consider a photonic hyper-crystal geometry which is based on hyperbolic material resonators separated by thin nonlinear dielectric layers, as shown in Fig.1. The hyperbolic resonators may be made of such artificial hyperbolic metamaterial as porous alumina ($Al_2O_3$) filled with silver nanowires, which exhibits low loss type I hyperbolic behavior ($\varepsilon_z<0$, $\varepsilon_x=\varepsilon_y>0$) in the visible range [9]. We should also note that pure $Al_2O_3$ itself exhibits low loss (Q>5) type I hyperbolic behavior in the 19.0-20.4 μm and 23-25 μm ranges [10]. The use of natural hyperbolic materials would greatly facilitate technical challenges involved in fabrication of the photonic hyper-crystal structures. While more technically challenging, quantum nonlinear optics in the LWIR range has been made possible due to recent introduction of the LWIR single photon detectors [11], so that consideration of natural hyperbolic materials would also appear to be justified. While propagation losses present a considerable challenge in the hyperbolic metamaterial design, Ni *et al.* [7] demonstrated that losses in a silver-based hyperbolic metamaterial may be compensated in the visible (~700 nm) frequency range by gain media, such as a dielectric polymer doped with Rh800 dye. When the dye is saturated, the silver-gain medium metamaterial structure becomes almost loss-free. One potential solution in the present case may consist in implanting the porous alumina with a suitable dye operating around 660 nm wavelength.

The hyperbolic resonators are assumed to be separated by thin nonlinear dielectric layers, such as lithium niobate as shown in Fig.1. Alternatively, if operation in the LWIR range is desirable, such nonlinear material as ZnSe (which is highly transparent in the LWIR range) may be used. A grating may be "buried" inside the nonlinear dielectric layer (as shown in Fig.1(b)) to enable controlled nonlinear photon-photon interaction. Another potentially interesting choice of the nonlinear layer would be to engineer a photonic hyper-crystal cavity and make use of the electromagnetically induced transparency (EIT) effects in either atomic impurities or quantum dots, as it is typically done in the conventional photonic crystal cavity geometries [2]. Kerr-like nonlinearities are enhanced by many orders of magnitude in such configurations. However, for the sake of simplicity let us initially assume that lithium niobate layers are used.

Let us determine the photon eigenfunctions in such a nonlinear photonic hyper-crystal geometry. In order to simplify our analysis, let us assume initially that $\varepsilon_{xy}=\varepsilon_1$ is positive and constant everywhere inside the hyper-crystal. Such an assumption is justifiable if similar to [9] the photonic hyper-crystal structure is operated at 660 nm, where the porous alumina/silver nanowire samples were measured to have $\varepsilon_x=\varepsilon_y=\varepsilon_1$ =5.28 and $\varepsilon_z=\varepsilon_2$ = - 4.8. It appears that $\varepsilon_1^{1/2}$ =2.3 indeed nearly matches the refractive index of lithium niobate [12]. Thus, at 660 nm wavelength $\varepsilon_z=\varepsilon_2$ will exhibit periodic oscillations inside the hyper-crystal as a function of z with a period $L<<\lambda$, while $\varepsilon_1$ may be assumed to be approximately constant. Far from the interfaces, inside the hyperbolic resonators $\varepsilon_2$= - 4.8=*const*, while near the $LiNbO_3$ interfaces $\varepsilon_z(z)$ experiences fast transition to its +5.28 value within the thin $LiNbO_3$ layers.

First, let us consider the macroscopic Maxwell equations inside the hyperbolic resonators far from the interfaces. The uniaxial symmetry of this medium reduces the ordinary and the extraordinary waves to respectively the TE ($\vec{E}\perp\hat{z}$) and TM ($\vec{B}\perp\hat{z}$) polarized modes. Let us introduce an extraordinary (TM) photon wave function as $\varphi=E_z$. The macroscopic Maxwell equations can be written as

$$\frac{\omega^2}{c^2}\vec{D}_\omega = \vec{\nabla}\times\vec{\nabla}\times\vec{E}_\omega \quad \text{and} \quad \vec{D}_\omega = \vec{\vec{\varepsilon}}_\omega \vec{E}_\omega , \tag{1}$$

which results in the following wave equation for the $\varphi_\omega$ field:

$$-\frac{1}{\varepsilon_1}\frac{\partial^2 \varphi_\omega}{\partial z^2} + \frac{1}{|\varepsilon_2|}\left(\frac{\partial^2 \varphi_\omega}{\partial x^2} + \frac{\partial^2 \varphi_\omega}{\partial y^2}\right) = \frac{\omega_0^2}{c^2}\varphi_\omega \tag{2}$$

Equation (2) is similar to the 3D Klein-Gordon equation describing a massive field propagating in a flat 2+1 dimensional Minkowski spacetime [13], in which the spatial z coordinate behaves as a timelike variable. The opposite signs of $\varepsilon_1$ and $\varepsilon_2$ lead to two important consequences. The dispersion law of the extraordinary waves in such a uniaxial material

$$\frac{k_x^2 + k_y^2}{\varepsilon_2} + \frac{k_z^2}{\varepsilon_1} = \frac{\omega^2}{c^2} \tag{3}$$

describes a hyperboloid in the phase space. As a result, the absolute value of the k-vector is not limited, and the volume of phase space between two such hyperboloids (corresponding to different values of frequency) is infinite. This divergence leads to a formally infinite (in the continuous medium limit) density of photonic states in the hyperbolic frequency bands of the metamaterial [13]. In the potential optical computing applications of nonlinear photonic hyper-crystals, optical bits in the geometry shown in Fig.1(a,b) may be encoded either via light intensity, or via the $k_{xy}$ component of the photon k vector, as indicated by the red and yellow arrows (the choice of optical bit encoding via the propagation direction is also quite common in optical computing geometries [1]). The buried grating which can be switched on and off via the nonlinear optical interactions of photons, mixes the photon states, thus enabling optical computations in the latter scheme.

Let us now consider the field behaviour near the periodic $Al_2O_3$/$LiNbO_3$ interfaces. Taking into account the translational symmetry of the system in $x$ and $y$ directions, we can still use the in-plane wave vector ($k_x$, $k_y$) as good quantum numbers, so that the propagating waves can be expressed as

$$E_\omega(\vec{r}) = E(z)\exp(ik_x x + ik_y y) \tag{4}$$

$$D_\omega(\vec{r}) = D(z)\exp(ik_x x + ik_y y)$$

$$B_\omega(\vec{r}) = B(z)\exp(ik_x x + ik_y y)$$

Because of the z dependence of $\varepsilon_z$, it is now more convenient to introduce the wavefunction $\psi(\vec{r})$ as the $z$-component of the electric displacement field of the TM wave:

$$\psi(\vec{r}) = D_z(\vec{r}) = \varepsilon_z(z)E_z(\vec{r}) = -\frac{c}{\omega}k_x B , \tag{5}$$

so that for the wave equation we obtain

$$-\frac{\partial^2 \psi}{\partial z^2} + \frac{\varepsilon_1}{\varepsilon_z(z)}\psi = \varepsilon_1 \frac{\omega^2}{c^2}\psi \tag{6}$$

In this wave equation the periodic $\varepsilon_1/\varepsilon_z$ ratio acts as a periodic effective potential. Solutions of eq.(6) may be found as Bloch waves

$$\psi(z) = \sum_{m=0}^{\infty} \psi_m \exp(i(k_z + \frac{2\pi}{L}m)z) \tag{7}$$

where $k_z$ is defined within the first Brillouin zone $-\pi/L < k_z < \pi/L$. Strong Bragg scattering is observed near the Brillouin zone boundaries at $k_z \sim \pi/L \gg \pi/\lambda$, leading to the formation of photonic bandgaps in both the wavenumber and the frequency domains. An example of the band structure calculations and a Bloch mode profile for a linear photonic hyper-crystal may be found in [3].

Let us consider this wave function behavior near the $Al_2O_3/LiNbO_3$ interfaces. Let us assume that the thickness of a very thin transition layer between $Al_2O_3$ and $LiNbO_3$ is very small but finite ($\delta \ll \lambda$), so that similar to [6], the following transition behaviour may be assumed near one of these interfaces located at z=0:

$$\varepsilon_Z(z) = \varepsilon_1 \varepsilon_2 \frac{(1-\exp(z/\delta))}{(\varepsilon_1 - \varepsilon_2 \exp(z/\delta))} + \frac{i\Gamma}{(1-(\varepsilon_2/\varepsilon_1)\exp(z/\delta))}, \tag{8}$$

where $\Gamma = 0.24$ equals the imaginary part of $\varepsilon_2$ of the wire array metamaterial at $\lambda=660$ nm [9] (a justification for such a transition layer to exist at the $Al_2O_3/LiNbO_3$ interface and extension of this model to the nonlinear optical properties of the interface will be done in Section 3). The corresponding effective potential $V=\varepsilon_1/\varepsilon_z$ experienced by extraordinary photons inside the photonic hyper-crystal is shown in Fig.2(a). Note that the potential step at the $Al_2O_3/LiNbO_3$ interface diverges in the limit of perfect loss compensation $\Gamma \to 0$ (see also Fig.3(a)). Substituting Eq.(8) into Eq.(6) and assuming $\Gamma \ll 1$, the wave equation near the $Al_2O_3/LiNbO_3$ interface may be re-written as

$$(u^2 + u)\frac{\partial^2 \psi}{\partial u^2} - \frac{Au + B}{u-1}\psi = 0, \tag{9}$$

where $u=\exp(z/\delta)$,

$$A = \left(k^2 - \frac{\varepsilon_1 \omega^2}{c^2}\right)\delta^2, \tag{10}$$

$$B = \left(\frac{\varepsilon_1 \omega^2}{c^2} - \frac{\varepsilon_1}{\varepsilon_2}k^2\right)\delta^2, \tag{11}$$

and $k$ is the wave number. Note that A<0 if the dielectric may support a propagating wave with wave number $k$, and A>0 otherwise. As described in detail in [6], the general solution of Eq. (9) in the limit $\Gamma \ll 1$ may be expressed in terms of the hypergeometric function $_2F_1(a,b,c,u)$ [14]:

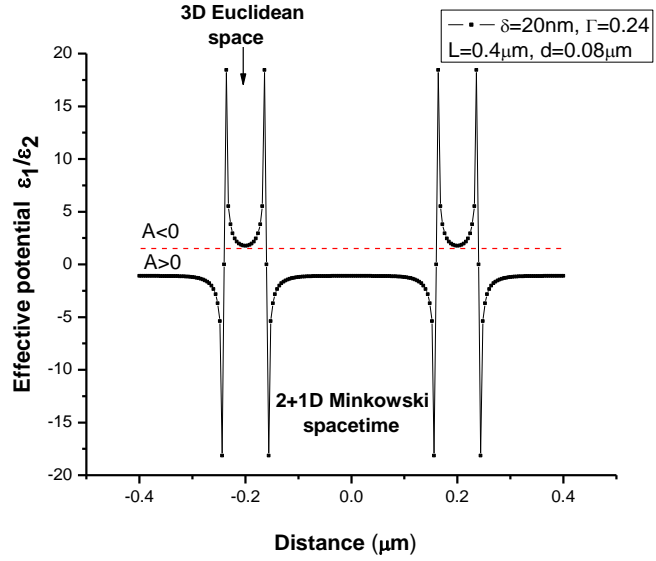

(a)

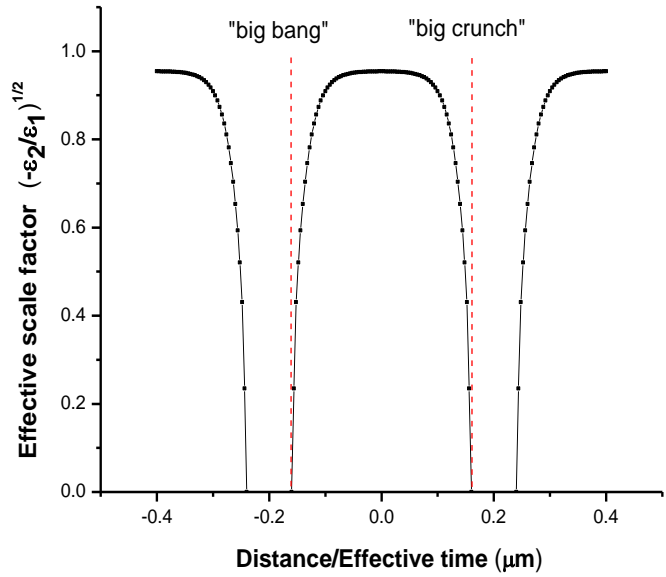

(b)

Fig. 2. (a) Effective potential $\varepsilon_1/\varepsilon_z$ experienced by extraordinary photons inside the Al$_2$O$_3$/LiNbO$_3$ photonic hypercrystal. In this simulation the hypercrystal parameters are: $L$=0.4μm, LiNbO$_3$ layer thickness is $d$=0.08μm, the transition layer thickness is $\delta$=20nm, $\Gamma$=0.24. (b) Scale factor $(\varepsilon_z/\varepsilon_1)^{1/2}$ of the effective optical 2+1D Minkowski spacetime experienced by the extraordinary photons inside the hyperbolic metamaterial resonators.

$$\psi(u) = u^{i\sqrt{B}} {}_2F_1^*(-\sqrt{A} - i\sqrt{B}, \sqrt{A} - i\sqrt{B}, 1 - 2i\sqrt{B}, u) +$$
$$+ ru^{-i\sqrt{B}} {}_2F_1^*(-\sqrt{A} + i\sqrt{B}, \sqrt{A} + i\sqrt{B}, 1 + 2i\sqrt{B}, u) \quad (12)$$

where the reflection coefficient $r$ is defined as

$$r = -\left[\frac{\Gamma(\sqrt{A} + i\sqrt{B})\Gamma(1 + \sqrt{A} + i\sqrt{B})}{\Gamma(\sqrt{A} - i\sqrt{B})\Gamma(1 + \sqrt{A} - i\sqrt{B})}\right]^* \frac{\Gamma(1 + 2i\sqrt{B})^*}{\Gamma(1 + 2i\sqrt{B})} \exp(-2\pi\sqrt{B}) \quad (13)$$

An example of calculated electric and magnetic field intensities in the $\Gamma \ll 1$ limit plotted near the interface in the case of A=30 and B=40 is shown in Fig.3(b). The electric field of extraordinary waves is strongly enhanced near the interface. For a wire array hyperbolic medium this field divergence may be explained via the well-known lightning rod effect at the tips of the silver nanowires.

We should also note that based on Eq.(2), the factor $(-\varepsilon_2/\varepsilon_1)^{1/2}$ plays the role of a scale factor of a 2+1 dimensional effective "optical spacetime", which "interval" may be introduced as

$$ds^2 = -dz^2 + (-\varepsilon_2/\varepsilon_1)(dx^2 + dy^2) \quad (14)$$

Since $\varepsilon_2$ is negative, in the case of hyperbolic metamaterials the concept of "optical spacetime" replaces the "optical space", which is typically introduced in transformation optics [15]. The scale factor of the effective Minkowski spacetime calculated using Fig.2(a) is plotted in Fig.2(b). If the spacetime analogy is used, each $Al_2O_3$/ $LiNbO_3$ interface of the photonic hyper-crystal behaves as either big bang or big crunch singularity. Inflation-like behaviour of the effective scale factor is observed near each interface, leading to extremely large field and strong nonlinearities near the interfaces.

## 3. Nonlinear optical interaction in hyperbolic resonator-based photonic hyper-crystals

Let us now consider how the introduction of photons into the individual hyperbolic resonators alters the effective potential $V = \varepsilon_1/\varepsilon_2$. The Maxwell-Garnett approximation may be used to evaluate the nonlinear optical effects in a wire array hyperbolic metamaterial structure and in the transition layer of thickness $\delta$ between the metamaterial and the lithium niobate. The diagonal components of the permittivity tensor of the wire array metamaterial may be obtained as [16]:

$$\varepsilon_1 = \varepsilon_{x,y} = \frac{2\alpha\varepsilon_m\varepsilon_d + (1-\alpha)\varepsilon_d(\varepsilon_d + \varepsilon_m)}{(1-\alpha)(\varepsilon_d + \varepsilon_m) + 2\alpha\varepsilon_d} \approx \frac{1+\alpha}{1-\alpha}\varepsilon_d \text{, and} \quad (15)$$

$$\varepsilon_2 = \varepsilon_z = \alpha\varepsilon_m + (1-\alpha)\varepsilon_d \quad (16)$$

where $\alpha$ is the volume fraction of the metallic phase, and $\varepsilon_m < 0$ and $\varepsilon_d > 0$ are the dielectric permittivities of the metal and dielectric component of the metamaterial, respectively (typically, $-\varepsilon_m \gg \varepsilon_d$). Assuming the same material parameters as in ref. [9] at 660 nm, the dielectric permittivity of $Al_2O_3$ is $\varepsilon_d = 2.4$, while the permittivity of silver is $\varepsilon_m = -21.6 + 0.8i$, so that $\alpha = 0.3$. Based on Eqs.(15,16), the transition layer of thickness $\delta$ between the metamaterial and the lithium niobate (which was approximated by Eq.(8) in Section 2) may be ascribed to the gradual $\alpha \to 0$ transition near the metamaterial surface, which is accompanied by surface roughness of the $Al_2O_3/LiNbO_3$ interface. We will assume that the nonlinear optical effects do not affect $\varepsilon_m$ (since light does not penetrate substantially into silver nanowires), so that only the dielectric permittivities of $Al_2O_3$ and $LiNbO_3$ will be influenced by the interfacial electric field divergence (shown in Fig.3(b)) via the nonlinear Kerr effect:

$$n = n_0 + n_2 I \qquad (17)$$

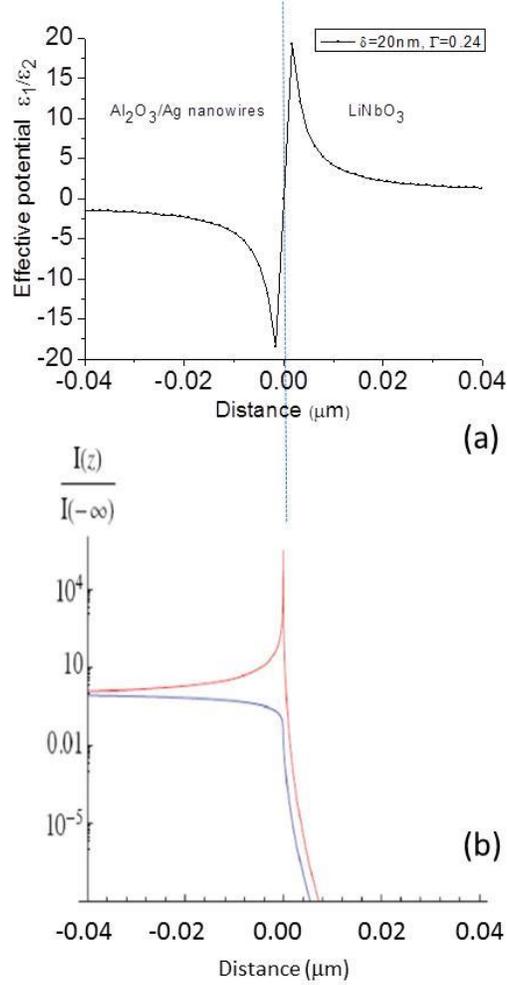

Fig. 3. (a) Effective potential $\varepsilon_1/\varepsilon z$ near the Al2O3/ LiNbO3 interface calculated without loss compensation using $\Gamma=0.24$ magnitude of the imaginary part of $\varepsilon_2$ of the wire array metamaterial [9]. Note that the potential step at the interface diverges in the limit of perfect loss compensation $\Gamma \to 0$. (b) Electric (red) and magnetic (blue) field intensities calculated near the interface in the $\Gamma<<1$ limit for the A>0 case (A=30, B=40).

According to [12], the nonlinear refractive index of doped $LiNbO_3$ reaches up to $n_{2+}=1.73 \times 10^{-10}$ cm$^2$/W, while much lower values $n_{2-} \sim 3.3 \times 10^{-16}$ cm$^2$/W of the nonlinear refractive index has been reported in the literature for $Al_2O_3$ [17]. Thus, based on the

arguments above and similar to Section 2, we may assume that near the interface located at z=0

$$n_2(z) = \frac{(n_{2+}+n_{2-})}{2} - \frac{(n_{2+}-n_{2-})}{2}\frac{(1-\exp(z/\delta))}{(1+\exp(z/\delta))} \approx \frac{n_{2+}\exp(z/\delta)}{(1+\exp(z/\delta))}, \quad (18)$$

and that the effective potential in the presence of photon field may be estimated as follows:

$$V(z) = \frac{\varepsilon_1}{\varepsilon_2} \approx \frac{\varepsilon_1^{(0)} + 2\sqrt{\varepsilon_1^{(0)}}n_2(z)I(z)}{\varepsilon_2^{(0)} + 2\sqrt{\varepsilon_1^{(0)}}n_2(z)I(z)} \approx \frac{\varepsilon_1^{(0)}}{\varepsilon_2^{(0)} + 2\sqrt{\varepsilon_1^{(0)}}n_2(z)I(z)}, \quad (19)$$

where $\varepsilon_1^{(0)}$ and $\varepsilon_2^{(0)}$ are the dielectric tensor components at $I$=0. Eq.(19) makes clear that the main effect of nonlinearities consists in shifting the position of the potential wall with respect to z=0. This behaviour is illustrated in Fig. 4(a) for two increasing levels of light intensity assuming 0.4x0.4x0.4 μm$^3$ hyperbolic resonator dimensions (same as in Fig.2) and either 2.3x10$^4$ or 7x10$^4$ photons in the resonator. Reducing the resonator dimensions to $\lambda/10$ (or 66 nm on the side) would reduce the required number of photons in the resonator to achieve the same effect to either 100 or 300, respectively. While this light level is relatively far from the single photon limit, Fig.4(b) illustrates that nonlinear photonic hyper-crystals based on such simple nonlinear dielectrics as LiNbO$_3$ or ZnSe may be used in classical optical computing schemes. For example, control light may be used to drastically change transmission of the photonic hyper-crystal via broadening of the tunnelling gaps around the nonlinear dielectric layers, as shown in Fig.5. Alternatively, in the bit encoding scheme based on the $k_{xy}$ direction, mode coupling properties of a buried grating may be altered between the "on" and "off" states via positioning the potential wall of the $V=\varepsilon_1/\varepsilon_2$ potential barrier either in front or behind the grating, as illustrated in Fig. 4(b). Both bit encoding schemes and both nonlinear interaction mechanisms may be used to realize classical optical gates, which may be operated using control light. As demonstrated e.g. in recent ref.[18], classical optical computing remains an attractive intermediate option between the conventional electronic classical computing and the future quantum computing schemes. Therefore, the optical "hyper-computing" scheme described in more detail in Section 4 below, may turn out to be useful even at the classical level.

An obvious way to increase $n_2$ by two to three orders of magnitude compared to LiNbO$_3$, and reach the single photon level is to implement a photonic hyper-crystal cavity design, and make use of the electromagnetically induced transparency (EIT) effects in either atomic impurities or quantum dots, as it is typically done in the conventional photonic crystal cavity geometries [2]. Kerr-like nonlinearities are enhanced by several orders of magnitude in such configurations, so that considerable photon-photon interaction may be observed at a single-photon level. Engineering of such photonic hyper-crystal nanocavities is possible due to formation of photonic band structure in hyperbolic metamaterials at any desired hyper-crystal periodicity $L$, which may be much smaller than the free space wavelength $\lambda$. Examples of the photonic hyper-crystal band structure calculations may be found in refs.[3,4]. Much higher levels of $n_2$ nonlinearity are achieved due considerably smaller mode volumes in such nanocavities. We should also mention a very recent study of Purcell enhancement of parametric luminescence in hyperbolic metamaterials [19], which reported a 1000 fold enhancement of the down-converted emission rate in experimentally realistic nanostructures, which gives yet another example of several orders of magnitude enhancement of nonlinear optical effects in hyperbolic nanostructures. In the next section we analyze potential applications of the nonlinear photonic hyper-crystals in optical limiting and optical computing devices.

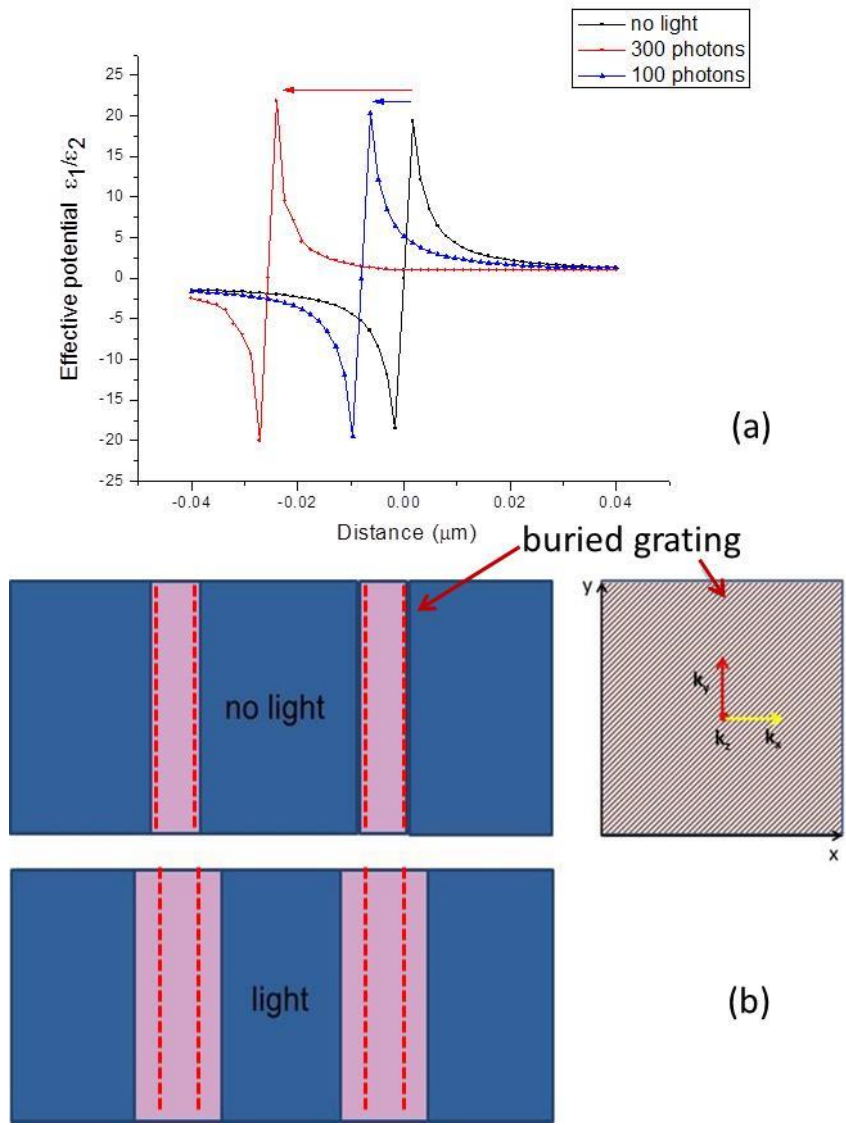

Fig. 4. (a) Shifting position of the potential wall with respect to z=0 using control light. These simulations assume 0.4x0.4x0.4 μm³ hyperbolic resonator dimensions (same as in Fig.2) and either 2.3x10⁴ or 7x10⁴ photons in the resonator. Reducing the resonator dimensions to $\lambda/10$ (or 66 nm on the side) would reduce the required number of photons in the resonator to achieve the same effect to approximately 100 and 300, respectively. (b) In a classical optical computing scheme control light may be used to drastically change transmission of the photonic hyper-crystal via broadening of the tunnelling gaps around the nonlinear dielectric layers. Alternatively, mode coupling properties of a buried grating may be altered between the "on" and "off" states via positioning the potential wall of the $V=\varepsilon_1/\varepsilon_2$ potential barrier either in front or behind the grating.

## 4. Potential applications of nonlinear photonic hyper-crystals in optical limiting and optical computing devices

Optical limiting for sensor and eye protection is an important application of nonlinear optics. However, progress in this area has been slow because of the limited range of fast nonlinear susceptibilities available in conventional optical materials. Recent development of nonlinear epsilon near zero (ENZ) metamaterials may offer a potential solution of this problem because of their unprecedented nonlinear response: the relative change Δn/n in refractive index becomes large as the permittivity becomes small, suggesting that near the ENZ frequencies a metamaterial system should exhibit extremely strong nonlinear optical properties. Unfortunately, the ENZ frequency range is typically very narrow in conventional metamaterials relying on resonant ENZ behavior (such as in conductive oxide-based metamaterials near their plasma frequency).

As we have seen from the previous discussion, similar to conventional hyperbolic materials, photonic hyper-crystals exhibit broadband epsilon near zero (ENZ) behaviour. In addition, due to the photonic crystal effects and the absence of diffraction limit, photonic hyper-crystals exhibit light localization on deep subwavelength scales, leading to giant enhancement of nonlinear optical effects. Compared to the other optical limiting solutions, nonlinear photonic hyper-crystals may provide considerable improvements in this application. As illustrated in Fig.5, in one of the potential optical limiting schemes intense control light may drastically change transmission of the photonic hyper-crystal via broadening of the effective tunnelling barriers (see Fig.2(a)) around the nonlinear dielectric layers. Tunnelling mechanism is an extremely (exponentially) sensitive way to suppress transmission of the external laser light by the optical limiter. In combination with the broadband ENZ behaviour of the hyperbolic metamaterials it may provide considerable improvements in the optical limiter performance.

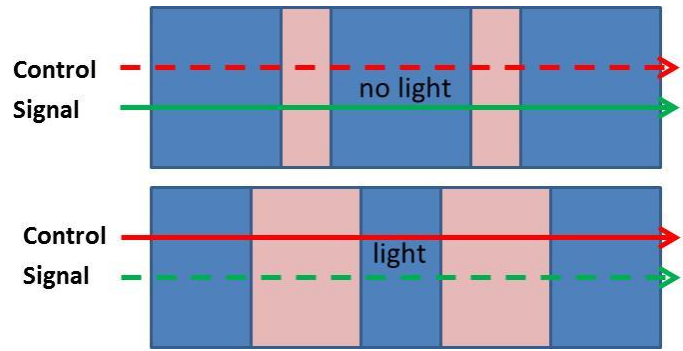

Fig. 5. In the optical limiting scheme intense light drastically changes transmission of the photonic hyper-crystal via broadening of the effective tunnelling barriers around the nonlinear dielectric layers (compare with Fig.4(a)).

Now let us discuss how the nonlinear optical effects described in the previous section may be used to build simple optical gates. As described in Section 2, optical bits in the geometry shown in Fig.1(a,b) may be encoded via the $k_{xy}$ component of the photon k vector, as indicated by the red and yellow arrows. This choice of optical bit encoding via the propagation direction is quite common in optical computing geometries [1]. The buried grating oriented at 45° with respect to the x and y axes, which can be switched on and off using the control light (as described in the previous section), mixes these optical bits, thus enabling realization of simple optical gates. As illustrated in Fig.4(b), in the absence of control light the 45° grating embedded into the nonlinear layer is exposed, and therefore it mixes the "0" and "1" states in equal proportions. Under the illumination with control light the effective "potential walls" move beyond the grating, so that the optical bit mixing is switched off. This simple example illustrates the viability of the photonic hyper-crystals based approach in optical computing schemes.

Let us now turn our attention to other non-trivial properties of nonlinear photonic hyper-crystals. Quantum optics of light in a hyperbolic metamaterial has been considered in detail in ref.[20]. The transition from classical to quantum optics occurs when the number of photons in any given mode is no longer large. Using the correspondence principle, the wave equation describing extraordinary photons propagating inside the hyperbolic metamaterial (Eq.(2)) may be re-written as follows:

$$\left(-\frac{1}{\varepsilon_1}\frac{\partial^2}{\partial z^2}+\frac{1}{|\varepsilon_2|}\left(\frac{\partial^2}{\partial x^2}+\frac{\partial^2}{\partial y^2}\right)-\frac{m^{*2}c^2}{\hbar^2}\right)\varphi_\omega=0 \quad (20)$$

where $\varphi_\omega$ is now understood as a quantum mechanical photon wave function, and the effective mass m* equals

$$m^*=\hbar\omega/c^2 \quad (21)$$

In the "non-relativistic" limit in which the kinetic energy (second term) in the parenthesis in Eq.(20) is much smaller than the effective rest energy $m^*c^2$, eq.(20) reduces to a standard Schrödinger equation:

$$i\frac{\hbar c}{\varepsilon_1}\frac{\partial}{\partial z}\varphi_\omega=\hat{H}\varphi_\omega=\pm\left(m^*c^2-\frac{\hbar^2}{|\varepsilon_2|2m^*}\left(\frac{\partial^2}{\partial x^2}+\frac{\partial^2}{\partial y^2}\right)\right)\varphi_\omega=\pm\left(m^*c^2+\frac{(\hat{p}_x^2+\hat{p}_y^2)}{|\varepsilon_2|2m^*}\right)\varphi_\omega \quad (22)$$

However, the role of effective Hamiltonian operator in this equation is played by

$$\hat{H}=i\frac{\hbar c}{\varepsilon_1}\frac{\partial}{\partial z} \quad (23)$$

(note that in the non-relativistic quantum mechanics the $m^*c^2$ term is usually omitted by re-defining the zero energy). If $\varepsilon_1$ and $\varepsilon_2$ are allowed to vary, an effective potential energy term would also appear in eq.(22). Thus, Eq.(22) replicates a version of non-relativistic quantum mechanics in a 2+1 dimensional Minkowski spacetime, in which the spatial *z* coordinate plays the role of effective time. Note also that in a more general "relativistic" situation where the effective kinetic energy is no longer much smaller than the rest energy $m^*c^2$, the "relativistic" Eq.(20) must be used. Thus, for all practical purposes the quantum optics of hyperbolic metamaterials and photonic hyper-crystals looks very similar to conventional quantum optics, except that the spatial z coordinate plays the role of time.

This property of hyperbolic metamaterials has a very interesting consequence. The speed of conventional computations (both classical and quantum) is ultimately limited by the speed of light. It limits, for example, the clock frequency of the computer. Since both classical and quantum computing rely on the clocks to perform a set of programmable steps, one step after another, a given computation requires a time interval given by the number of computational steps *N* divided by the clock frequency *ν*. The model of quantum optics outlined above seems to be able to map such a conventional (temporal) computation onto a computation performed in a hyperbolic metamaterial using *N* spatial steps instead of *N* temporal ones. As illustrated in Fig.6, due to such a periodic spatial "clock", which is used in the mapping, the hyperbolic metamaterial becomes a photonic hyper-crystal. The top portion of the figure shows schematically the succession of NOR operations performed on a set of several bits in a conventional (temporal) computer. These operations are performed one at a time in sync with a temporal clock signal. The bottom portion of Fig.6 shows how these operations may be mapped onto operations performed with optical bits in a photonic hyper-crystal using the optical limiting effect, where A and B rays play the role of control and weaker X ray acts as an output signal. The hyper-crystal periodicity plays the role of the clock signal. The potential advantage of such a photonic hyper-crystal based hyper-computing scheme would be a much faster computation speed, which is important in time-sensitive applications.

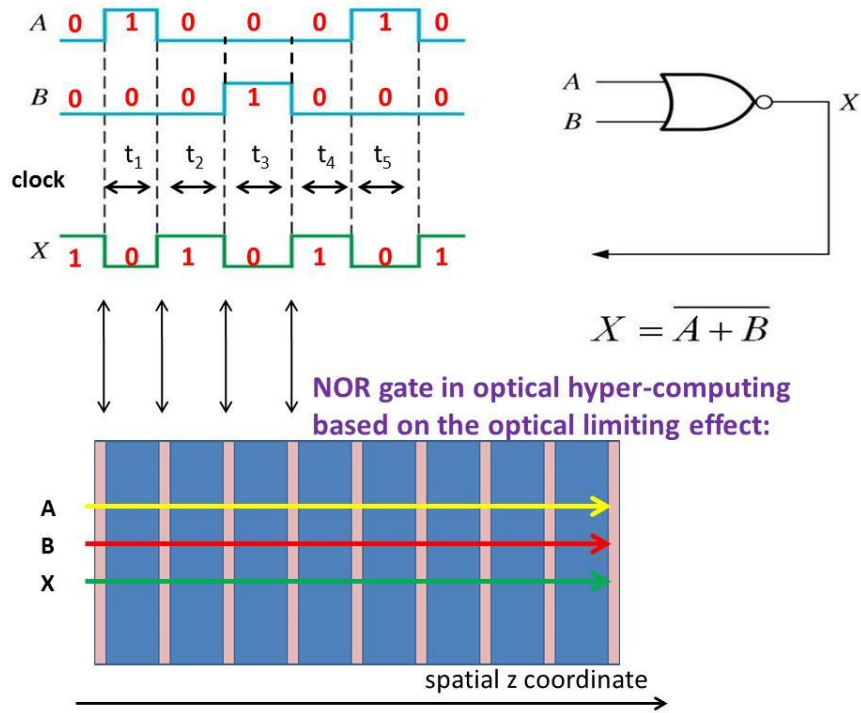
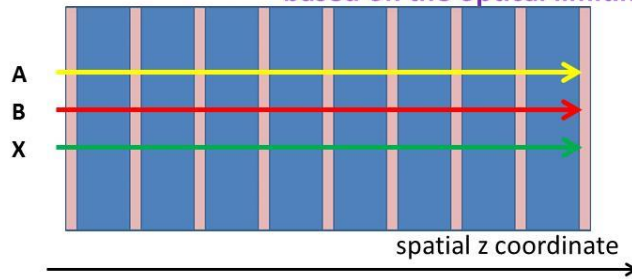

Fig. 6. Mapping of a conventional (temporal) computation onto an optical computation performed in a hyperbolic metamaterial using N spatial steps instead of N temporal ones: The top portion of the figure shows schematically the succession of NOR operations performed on a set of several bits in a conventional (temporal) computer. These operations are performed one at a time in sync with a temporal clock signal. The bottom portion of this figure shows how these operations may be mapped onto operations performed with optical bits in a photonic hyper-crystal using the optical limiting effect, where A and B rays play the role of control and weaker X ray acts as an output signal. The hyper-crystal periodicity plays the role of the clock signal.

We should also note that the optical hyper-computing proposal outlined above have some features in common with the recently suggested "faster-than-light" computing proposal based on immersion of an optical computer into a medium having index of refraction smaller than one [21], thereby trespassing the speed-of-light communication barrier. While both "faster-than-light" and "hyper-computing" proposals rely on the artificial metamaterial media for their accelerated operation, the proposal of Putz and Svozil [21] still relies on the more conventional "temporal" computing and communication scheme.

The large density of optical states in hyperbolic metamaterials implies reduced group velocity compared to the velocity of light in vacuum, which may impede the optical hyper-computing computation scheme. However, the magnitude of group velocity remains very large in hyperbolic materials. For example, recent detailed calculations performed in [22] indicate group velocities in the hyperbolic frequency domains of realistic metal-dielectric metamaterials of the order of $v_g \sim 0.1c_0$-$0.3c_0$, where $c_0$ is the velocity of light in vacuum. Since the photonic hyper-crystal geometry implies subwavelength dimensions $L$ of the hyperbolic metamaterial resonators (see Fig.1a), the corresponding information propagation time between the resonators $L/v_g$ would fall into the femtosecond range, which far exceeds the ~GHz clock speed of both classical and quantum state of the art computers.

Another potential limitation to the suggested optical hyper-computation scheme may be presented by the need to find a right balance between the propagation losses and the potential back-scattering issues in the proposed nonlinear photonic hyper-crystals. Too much propagation loss will make the individual optical bits non-interacting, while too little loss may lead to undesirable back-scattering effects. The proper balance between these opposite effects may be found by properly choosing the gain and the length of the individual hyperbolic resonators in the photonic hyper-crystal structure.

## 5. Conclusion

In conclusion, we have demonstrated that photonic hyper-crystals exhibit light localization on deep subwavelength scales, leading to considerable enhancement of nonlinear photon-photon interaction. Like their conventional photonic crystal counterparts, nonlinear photonic hyper-crystals appear to be very promising in optical limiting and optical computing applications. Since spatial coordinate plays the role of time in this geometry, an optical "hyper-computing" scheme may be suggested, which considerably accelerates computation time, leading to considerable advantages over the conventional computing schemes in time-sensitive applications.

**References**

1. J. L. O'Brien, "Optical quantum computing", Science **318**, 1567-1570 (2007).
2. H. Choi, M. Heuck, D. Englund, "Self-similar nanocavity design with ultrasmall mode volume for single-photon nonlinearities", Phys. Rev. Lett. **118**, 223605 (2017).
3. V. N. Smolyaninova, B. Yost, D. Lahneman, E. Narimanov, I. I. Smolyaninov, "Self-assembled tunable photonic hyper-crystals", Scientific Reports **4**, 5706 (2014).
4. E. E. Narimanov, "Photonic hypercrystals", Phys. Rev. X **4**, 041014 (2014).
5. I. I. Smolyaninov, "Analog of gravitational force in hyperbolic metamaterials", Phys. Rev. A **88**, 033843 (2013).
6. I. I. Smolyaninov, E. Hwang, E. E. Narimanov, "Hyperbolic metamaterial interfaces: Hawking radiation from Rindler horizons and spacetime signature transtions", Phys. Rev. B **85**, 235122 (2012).
7. X. Ni, S. Ishii, M. D. Thoreson, V. M. Shalaev, S. Han, S. Lee, A. V. Kildishev, "Loss-compensated and active hyperbolic metamaterials", Optics Express **19**, 25242-25254 (2011).
8. A. P. Slobozhanyuk, P. Ginzburg, D. A. Powell, I. Iorsh, A. S. Shalin, P. Segovia, A. V. Krasavin, G, A. Wurtz, V, A. Podolskiy, P, A. Belov, A. V. Zayats, "Purcell effect in hyperbolic metamaterial resonators", Phys. Rev. B **92**, 195127 (2015).
9. J. Yao, Z. Liu, Y. Liu, Y. Wang, C. Sun, G. Bartal, A. M. Stacy, X. Zhang, "Optical negative refraction in bulk metamaterials of nanowires", Science **321**, 930 (2008).
10. K. Korzeb, M. Gajc, D. A. Pawlak, "Compendium of natural hyperbolic materials", Optics Express **23**, 25406-25424 (2015).
11. T. Ueda, Z. An, K. Hirakawa, S. Komiyama, "Single photon detection in the long wave infrared", In: B. Murdin, S. Clowes (eds) Narrow Gap Semiconductors 2007. Springer Proceedings in Physics, vol 119. Springer, Dordrecht.
12. L. Palfalvi, J. Hebling, G. Almasi, A. Peter, K. Polgar, K. Lengyel, R. Szipocs, "Nonlinear refraction and absorption of Mg doped stoichiometric and congruent LiNbO3", J. Appl. Phys. **95**, 902-908 (2004).

13. I. I. Smolyaninov, E. E. Narimanov, "Metric signature transitions in optical metamaterials", Phys. Rev. Letters **105**, 067402 (2010).
14. G. N. Watson, A Treatise on the Theory of Bessel Functions (Cambridge University Press, Cambridge, 1944).
15. J. B. Pendry, D. Schurig, D. R. Smith, "Controlling electromagnetic fields", Science **312**, 1780 (2006).
16. R. Wangberg, J. Elser, E. E. Narimanov, and V. A. Podolskiy, "Nonmagnetic nanocomposites for optical and infrared negative-refractive-index media", J. Opt. Soc. Am. B **23**, 498-505 (2006).
17. A. Major, F. Yoshino, I. Nikolakakos, J. S. Aitchison, P. W. E. Smith, "Dispersion of the nonlinear refractive index in sapphire", Opt. Letters **29**, 602-604 (2004).
18. F. Beil, T. Halfmann, F. Remacle, R. D. Levine, "Logic operations in a doped solid driven by stimulated Raman adiabatic passage", Phys. Rev. A **83**, 033421 (2011).
19. A. Davoyan, H. A. Atwater, "Quantum nonlinear light emission in metamaterials: broadband Purcell enhancement of parametric downconversion", Optica **5**, 5 (2018).
20. I. I. Smolyaninov, "Quantum mechanics of hyperbolic metamaterials: Modeling of quantum time and Everett's universal wavefunction", Physica B **453**, 131-135 (2014).
21. V. Putz, K, Svozil, "Can a computer be "pushed" to perform faster-than-light?", arXiv:1003.1238 [physics.gen-ph], presented at the UC10 Hypercomputation Workshop "HyperNet 10", Tokyo, June 22, 2010.
22. P.-G. Luan, J.-L. Wu, "On the possibility of superluminal energy propagation in a hyperbolic metamaterial of metal-dielectric layers", AIP Advances **8**, 015106 (2018).